\documentstyle[12pt]{article} \topmargin=-0.4in

\def\be#1#2{\begin{equation}{#1}\label{{#2}}\end{equation}}

\begin{document} 
\begin{center}
{\Huge Curved, extended classical solutions}\\ 
\vskip0.2in
{\Large \bf  I. The undulating
kink }\\
\vskip0.1in
A. Herat, R. Rademacher, and P. Suranyi, \\{\em  Department of Physics, University
of Cincinnati, Cincinnati, Ohio 45221-0011.}
\end{center}
\vskip0.2in
\abstract{
The energy of extended classical objects, such as vortices, depends on their
shape.  In particular, we show that the curvature energy of a kink in two spatial
dimensions, as a prototype of extended classical solutions, is always negative.  We
obtain a closed form for the curvature energy, assuming small deviations from the
straight line. }
 \vskip0.2in PACS numbers:  11.27+d, 11-10.Lm, 03.70.+k

\section{Introduction}

Classical solutions play an important role in a host of applications.  Besides
instantons and monopoles, vortices, abelian and nonabelian, acquired  central
importance, among others, in superconductors~\cite{abrikosov}, cosmic
strings~\cite{kibble}, and as objects responsible for
confinement~\cite{nielsen}~\cite{mandelstam}. Classical solutions of  recent
interest are $p$-branes, appearing in string theories.~\cite{strings} 

Vortices and
some other solutions of field equations distinguish themselves by the fact
that they are extensions of finite energy classical solutions (solitons) into
one or more additional spatial dimensions. As such, they have infinite energy, but
finite energy per unit length, if the additional space dimensions are
infinite. The  energy of these solutions becomes finite if the additional spatial
dimensions are finite. In what follows we will only consider finite
additional dimensions that make analytic studies of these objects possible.

To the best of our knowledge, analytic studies of
these objects have not been extended to study some of their important properties.
One  aspect of classical, finite energy solutions lifted into higher
dimensions, is that their energy is dependent on their shape and length.  In most
applications it has been assumed~\cite{nelson},\cite{rebbi}  that the energy of a 
3-dimensional vortex is simply its length multiplied by the energy of unit length
(which is equal to the energy of the 2 dimensional soliton), irrespective of its
shape. 

Before proceeding any further we need to clarify  what we mean under the
dependence of the energy on shape.  Topologically stable classical solutions are
characterized by the vanishing of an order parameter (usually a scalar field) at a
finite coordinate, which is, by definition,  the location of the center or core of
the object.  In trivially extended classical solutions the zeros form a straight
line.  These extended objects are solutions of higher dimensional field equations. 
Suppose now we impose a constraint that the locus of zeros lies on a predetermined
curve and minimize the Hamiltonian with this constraint.  Then, in general, the
value of the Hamiltonian will depend on the shape and length of the locus of zeros.
This dependence is important, because before one can attempt an analytic or
numerical study of the interaction and condensation of vortices, or of other
classical extended objects, one needs to clarify the energy of a single object.  As
we will see later, the dependence of the energy on the length and curvature of an
undulating object  is highly nontrivial. 

One might think that the deviation of an extended object from the minimum energy
straight line is a purely quantum phenomenon and should be treated in perturbation
theory.  Such a treatment would not, however, adequately describe large scale
features of extended objects.  Such, macroscopic features of vortices are clearly
seen in superconductors or simulations of cosmic strings.\cite{rebbi} Therefore, it
should be possible to calculate the relevant properties, at least in some
approximation, studying the classical equations of motion. If the energy and
transverse distribution of curved vortices can be calculated then their contribution
to a functional integral, i.e. their full quantum contribution, can also be
evaluated, at least in  perturbation theory.  

One of the ultimate targets of our investigations are vortex solutions in gauge
theories. As no analytic vortex solutions exist, the problem of
curved vortices is fairly difficult. Though our aim is to understand
issues pertaining to the curvature dependence of classical solutions in more
complicated quantum theories, in the current paper we will discuss the curvature
energy for a much simpler problem, the kink solution of the 2+1 dimensional linear
sigma model.  We hope to gain an
understanding of this problem that can be usefully applied to the more difficult
problem of vortices and other extended classical objects. Vortices and other
extended objects will be investigated in subsequent publications.

\section{The kink solution in 1+1 dimension}

One of the simplest examples  of classical solutions is the one dimensional kink,
which is a soliton solution of the two dimensional linear sigma model. Kinks are
topologically stable finite energy solutions that can be extended to 2 dimensions in
a straightforward manner. We start with a brief description of these, well known
solutions,~\cite{raja} with the intention of establishing notations.

 As the soliton solution can be chosen to be static, the Lagrangian coincides
with the negative of the Hamiltonian. The  Hamiltonian for a time
independent solution is 
\begin{equation}
H=\frac{1}{2}\int d\xi \left[\left(\frac{\partial \Phi}{\partial
\xi}\right)^2+\frac{\lambda}{2}\left(\Phi^2-\frac{m^2}{\lambda}\right)^2\right].
\label{ham1d}
\end{equation}
The vacuum expectation value of field $\Phi$, $m^2/\lambda$, and the coupling
constant can be scaled out if we use the transformation
\begin{equation}
\Phi(\xi-\xi_0)=\frac{m}{\sqrt{\lambda}}\Psi(x-x_0),
\label{rescale}
\end{equation}
where $m(\xi-\xi_0)/\sqrt{2}=x-x_0$. The rescaled Hamiltonian takes the form
\begin{equation}
H=\frac{m^3}{2\sqrt{2}\lambda}\int dx \left[\left(\frac{\partial \Psi}{\partial
x}\right)^2+\left(\Psi^2-1\right)^2\right].
\label{ham1d2}
\end{equation}
(\ref{ham1d2}) is minimized by a kink (anti-kink) solution
\begin{equation}
\Psi_c(x-x_0) =\pm \tanh(x-x_0).
\label{kink}
\end{equation}
The value of the Hamiltonian at this  solution is 
\begin{equation}
H=H_0=\frac{m^3}{\sqrt{2}\lambda}\frac{4}{3}.
\label{optimum}
\end{equation}
In what follows, we will set $m^2=\lambda=2$. The correct units can be easily
restored.

\section{Extension to 2+1 dimensions}

In  the three dimensional linear sigma model there is an extra term in the
Hamiltonian density, $(\partial \Phi/\partial y)^2$. Solution (\ref{kink}) also
satisfies the field equation in 2 dimensions, but it would have  infinite energy if 
space is infinite in the
$y$ direction. As we indicated earlier, we choose a finite
$y$ dimension, with periodic boundary condition. Then the kink  energy is
finite.  In other words, we seek a kink solution on an infinite
strip of width $L$. 

The minimum energy kink on the strip is a straight line at a fixed value
of
$x$. It is interesting to study, however, the dependence of the energy on the shape
of the kink.   For this purpose we distort the shape of the
kink by assuming that the zero of the solution is not at
$x_0=0$ (this value is chosen arbitrarily) but rather at $x=\epsilon(y)$.    The
periodicity condition, with the translation symmetry along the $x$ axis allows us
to choose
$\epsilon(0)=\epsilon(L)=0$. 

The problem we pose is the following. Suppose a closed curve, defined by
$x=\epsilon(y)$,  winds through the strip.  This curve is the locus of the zeros
of
$\Phi(x,y)$. In other words, we impose the constraint on the field,
$\Phi(\epsilon(y),y)=0$. We wish to find the field configuration,
$\Phi(x,y)$, satisfying this constraint and minimizing the Hamiltonian.  We are
mostly interested in the dependence of the  minimum of the Hamiltonian on the shape
of the curve.

The two-dimensional Hamiltonian has the form 
\begin{equation}
H=\frac{1}{2}\int dx\,dy\,\left[\left(\frac{\partial\Phi}{\partial x}\right)^2 +
\left(\frac{\partial\Phi}{\partial y}\right)^2+(\Phi^2-1)^2\right].
\label{ham2d}
\end{equation}
We assume here that $\Phi$ is a configuration minimizing $H$ for a given shape of
the core.

Define the {\em length-energy,} $H_l$, of a two a dimensional curved kink solution
as
$H_l=l H_0$, where $l$ is the total length of the kink,\[
l=\int_0^L dy \sqrt{1+\epsilon'^2}.
\]
 and $H_0$ is the energy of
the one dimensional kink (energy of unit length of straight two dimensional kink). 
Then we define the {\em curvature energy} of a two dimensional kink solution by
subtracting the `length-energy,' from the total energy, $H_c=H-H_l$. 

We can then
 prove the following theorem:\\ {\em The curvature energy of the two dimensional
kink is non-positive.} \\ In fact, detailed calculations will show that the
curvature energy is negative definite.  In other words, discounting the length
energy, the kink prefers the curled state to the straight one.

Proof: Consider a class ${\cal C}$ of
solutions that depend on the coordinate $x$ and $y$ only through the combination 
$x-\epsilon(y)$.  The Hamiltonian minimized on functions of this class will have a
minimum that is not smaller than the unrestricted minimum. To prove the theorem we
will show now that the Hamiltonian minimized on ${\cal C}$ is exactly the length
energy. 

Assume that $\Phi$ depends on $x$ and $y$ through $x-\epsilon(y)$. Then upon
substitution into (\ref{ham2d}) and shift of coordinate $x$ by $\epsilon(y)$ we
obtain
\begin{equation}
H=\frac{1}{2}\int dx\,dy\,\left[(1+{\epsilon'}^2)\left(\frac{d\Phi}{dx}\right)^2 +
(\Phi^2-1)^2\right].
\label{ham2d2}
\end{equation}
The substitution $x\to x\sqrt{1+{\epsilon'}^2}$ leads to 
\begin{equation}
H=\frac{1}{2}\int
dx\,dy\,\sqrt{1+{\epsilon'}^2}\left[\left(\frac{d\Phi}{dx}\right)^2
+ (\Phi^2-1)^2\right]=lH_0.
\label{ham2d3}
\end{equation}
where 
\[
l=\int dy \sqrt{1+{\epsilon'}^2}.
\]
$\Phi$ depends on $x$ only.
As the one dimensional Hamiltonian is minimized by the original kink solution,
(\ref{kink}), we obtain $H=4l/3$, as we asserted above. This completes the proof of
the theorem.

\section{Exact solution for small deviations from the straight line}

In the previous section we showed that the curvature energy of an arbitrary
curved kink is negative.  We are able to calculate the curvature energy
analytically only if
 we assume that the deviation of the kink from the straight line,
$\epsilon(y)$, is small.  

The general form of the two dimensional kink solution can
always be written as
 $\Phi(x,y)=\Phi_c(x-\epsilon)+\chi(x-\epsilon,y)$,  where
 $\Phi_c=\tanh x$. As our constraint is $\Phi(\epsilon(y),y)=0$  and, having
$\Phi_c(0)=0$, we also have
\begin{equation}
\chi(0,y)=0.\label{condition}\end{equation}
Since at $\epsilon=0$ the solution $\Phi_c$ is exact, 
$\chi=O(\epsilon)$, as well.

Now we can vary the Hamiltonian with respect to $\chi$ with the subsidiary
condition (\ref{condition}). Keeping terms of $O(\epsilon)$ we obtain the following
equation of motion for $\chi$:
\begin{equation}
\chi_{xx}+\chi_{yy}-\epsilon''(y)\Phi_c'-2(3\Phi_c^2-1)\chi=\delta(x)\lambda(y),
\label{chi-eq}
\end{equation}
where the Lagrange multiplier $\lambda(y)=\chi_x(\epsilon+0,y)-\chi_x(\epsilon-0,y)$
and where the subscripts $x$ and $y$ indicate partial derivatives.
The appearance of the delta function in (\ref{chi-eq}) has no other significance in
our subsequent calculations than removing a condition  on the continuity of
the {\em derivative} of
$\chi$ at
$x=0$.  

Notice that the inhomogeneous driving term $\epsilon''(y)\Phi_c'$ in
(\ref{chi-eq}) is an even function of $x$.  Then due to the $x\to-x$ symmetry of
(\ref{chi-eq}) we also have
$\chi(x,y)=\chi(-x,y)$. It is sufficient to solve (\ref{chi-eq}) for
$x>0$, with  boundary conditions $\chi(0,y)=\chi(\infty,y)= 0$. Note that
though $\chi(x,y)$ is even $\chi_x(0,y)\not=0$ because it is not
continuous. 
 
To
find the appropriate solution we must  find   solutions of the homogeneous
 and   inhomogeneous equations that vanish at $x\to\infty$ and find a
combination of these two solutions that also vanishes at the origin.

Let us expand $\chi(x,y)$ in a Fourier series of $y$. We will write
\be{
\chi(x,y)=\frac{1}{\sqrt{L}}\sum_n e^{2\pi i\, n y/L}F_n(x),
}{expand}
where $F_{-n}=F_n^\star$.
Using $\Phi_c=\tanh x$ the modes satisfy the one-dimensional inhomogeneous
Schr\"odinger equation
\begin{equation}
F_n''+\frac{6}{\cosh^2x}F_n -
4\left(1+\frac{\pi^2n^2}{L^2}\right)F_n=-\frac{4\pi^2n^2}{L^2}\frac{\epsilon_n}{\cosh^2x}.
\label{modes}
\end{equation}
where $\epsilon_n$ is the Fourier expansion coefficient of $\epsilon(y)$. 

The solution of the inhomogeneous
Schr\"odinger equation satisfying boundary conditions $F_n(0)=0$ and $F_n(x)\to 0$
at
$x\to\infty$ is 
\begin{equation}
F_n=\epsilon_n\left[\frac{1}{\cosh^2x}-e^{-\alpha
x}\left(1+\frac{3\alpha}{\alpha^2-1}\tanh x
+\frac{3}{\alpha^2-1}\tanh^2x\right)\right].
\label{final}
\end{equation}
where 
\begin{equation}
\alpha=2\sqrt{1+
\frac{\pi^2n^2}{L^2}}.
\label{parameters}
\end{equation}

The $O(\epsilon^2)$ deviation of the Hamiltonian from that of a straight kink
is  \begin{equation}
\Delta H=\frac{1}{2}\int dx dy
\left[(\chi_x)^2+(\chi_y-\epsilon'\Phi_c')^2+2(3\Phi_c^2-1)\chi^2\right].\label{ham4}
\end{equation}
After integrating by parts and  utilizing (\ref{chi-eq}) we obtain
\begin{equation}
\Delta H=\frac{1}{2}\int dx dy\epsilon'\Phi'_c(\epsilon'\Phi'_c-\chi_{y})
\label{partial}
\end{equation}

Substituting solution (\ref{final}) into (\ref{partial}) the integral can be easily
evaluated to give the final result
\begin{equation}
\Delta H=\sum_n\epsilon_n^2\frac{4n^2\pi^2}{L^2} \frac{2}{\alpha(\alpha^2-1)}
=\sum_n\epsilon_n^2\frac{8n^2\pi^2}{L^2}\frac{\sqrt{1+\frac{\pi^2n^2}{L^2}}}
{
\frac{4\pi^2n^2}{L^2}+3}.
\label{complete}
\end{equation}

Compare (\ref{complete}) with (\ref{ham2d3}). The increase of length energy, as
obtained from the expansion of (\ref{ham2d3}) is
\begin{equation}
\Delta H_0   =H_0\Delta l\simeq\frac{H_0}{2}\int dy
[\epsilon'(y)]^2=\sum_n\epsilon_n^2\frac{8\pi^2n^2}{3L^2},
\label{ham8}
\end{equation}
where $\Delta l=l-L$ is the increase of the length of the kink due to undulation.
Then  comparing (\ref{ham8}) with (\ref{complete}) shows that $\Delta H_0>\Delta H$,
or in other words, the curvature energy is always negative. 

Assuming that long wavelength modes dominate we can expand the expression of the
excess energy in $\pi^2n^2/L^2$. Then using 
 \[
R=\frac{1+\epsilon'^2}{|\epsilon''|}\simeq \frac{1}{|\epsilon''|},
\]
where $R$ is the curvature radius
we get the leading order correction term in long wavelength modes 
\[
\Delta
H\simeq\sum_n\epsilon_n^2\left[\frac{8\pi^2n^2}{3L^2}-\frac{2\pi^4n^4}{3L^4}
+\dots\right]
=H_0\left(\Delta l-\frac{1}{32}\int\frac{ds}{R^2}+\dots\right). \]
Omitted 
higher order terms will contain derivatives of the curvature radius. The
minimum of the total Hamiltonian is
\[
H=H_0\left[\int ds-\frac{1}{32}\int ds( \mbox{\boldmath$ x$}'')^2+\dots\right] ,
\]
where \mbox{\boldmath$ x$} is the 2-dimensional coordinate vector of the
undulating kink core in pathlength gauge.  Clearly, the curvature term is negative,
as required by the theorem proven in the previous section.
\section{Summary and future work}

We have shown that the energy of a two dimensional extension of the one
dimensional kink solution is always smaller than the energy of the one
dimensional kink multiplied by the length of the two dimensional kink. 
Consequently, the curvature energy is always negative.  We found the exact
solution of the constrained two dimensional field equation using the approximation
that the deviation of the shape of the kink from the straight line is small. The low
frequency contributions to this energy can be interpreted as a positive length term
and a negative curvature term. 

We regard this work as an exercise preceeding the investigation of similar problems
for two dimensional solitons extended to three dimensional vortices in gauge
theories. In a follow-up paper we will use similar methods for finding the energy
of a curved vortex.  As no analytic solutions for vortex solutions exist we will
have to perform some of the calculations numerically. 
\\

\vskip0.2in
This work was supported in part by
 the U.S. Department of Energy through grant
\#DE FG02-84ER-40153.

   \end{document}